\documentclass[twocolumn,showpacs,preprintnumbers,amsmath,amssymb]{revtex4}
 \usepackage{graphicx}  

\newcommand{\beq}{\begin{eqnarray}}
\newcommand{\eeq}{\end{eqnarray}}

\begin{document}

\title{Multi-pomeron repulsion and the Neutron-star mass}

\author{Y.\ Yamamoto$^{1}$}
\author{T.\ Furumoto$^{2}$}
\author{N.\ Yasutake$^{3}$}
\author{Th.A.\ Rijken$^{4}$$^{1}$}
\affiliation{
$^{1}$Nishina Center for Accelerator-Based Science,
Institute for Physical and Chemical
Research (RIKEN), Wako, Saitama, 351-0198, Japan\\
$^{1}$Ichinoseki National College of Technology,
Ichinoseki, Iwate, 021-8511, Japan\\
$^{3}$Department of Physics, Chiba Institute of Technology, 2-1-1 Shibazono
Narashino, Chiba 275-0023, Japan\\
$^{4}$IMAPP, University of Nijmegen, Nijmegen, The Netherlands
}

%

\begin{abstract}
A multi-pomeron exchange potential (MPP) is proposed as a model for
the three-body repulsion indicated in neutron-star matter,
which works universally among three- and four-baryons.
Its strength is determined by analyzing the nucleus-nucleus scattering
with the G-matrix folding model. The EoS in neutron matter is
obtained including the MPP contribution. 
The neutron-star mass is calculated by solving the TOV equation.
The maximum mass is obtained to be larger than the observed one 
$1.97 M_{solar}$ on the basis of the experimental data.
\end{abstract}


\maketitle

\parindent 15 pt

\section{Introduction}

The new observation of the system J1614-2230~\cite{Demorest10}
has brought a great impact on the maximum-mass problem
of neutron stars.  The large observed value of neutron-star 
mass $1.97 M_{solar}$ gives a severe condition for the
stiffness of the equation of state (EoS) of neutron-star matter.

For realization of the nuclear saturation property 
based on underlying nuclear interactions,
an essential role is played by the three-nucleon ($N$) interaction (TNI)
composed of the attractive part (TNA) and the repulsive part (TNR).
Especially, the TNR contribution increasing rapidly in 
the high-density region leads to a high values of 
nuclear incompressibility:
The TNR contribution in high-density neutron matter plays an essential 
role for stiffening of the EoS of neutron-star matter, assuring the 
observed maximum mass of neutron stars.
%

However, the hyperon ($Y$) mixing in neutron-star matter brings about 
the remarkable softening of the EoS, which cancels the TNR effect
for the maximum mass. 
In order to avoid this serious problem, Nishizaki, Takatsuka 
and one of the authors (Y.Y.) \cite{NYT} introduced the conjecture that 
the TNR-type repulsions work universally for $Y\!N$ and $Y\!Y$ 
as well as for $N\!N$.  
They showed that the role of the TNR for stiffening the EoS
can be recovered clearly by this assumption. 
Our basic concern is the existence of universal repulsions among
three baryons, called here as the three-baryon repulsion (TBR).

In modeling for $Y\!N$ and $Y\!Y$ interactions,
important development has been accomplished 
by the Extended Soft Core (ESC) models.
Here, two-meson and meson-pair exchanges are taken into 
account explicitly and no effective boson is included differently 
from the usual one-boson exchange models. The latest version of 
ESC model is named as ESC08c~\cite{ESC08,ESC08c}.
Hereafter, ESC means this version.
In this work, TBR is taken into account by the 
multi-pomeron exchange potential (MPP) within the ESC modeling.
In order to reproduce the nuclear saturation property precisely,
it is necessary to introduce also TNA.
We treat this part phenomenologically.


The G-matrix theory gives a good starting point 
for studies of many-body systems on the basis of
free-space baryon-baryon interaction models.
Here, the correlations induced by short-range and tensor components
are renormalized into G-matrix interactions.
In the case of nucleon matter, the lowest-order G-matrix calculations 
with the continuous (CON) choice for intermediate single particle potentials
were shown to simulate well the results including higher hole-line contributions
up to $3\sim 4$ times of normal density $\rho_0$~\cite{Baldo02}.
On the basis of this recognition,
one can study properties of high-density nuclear matter
using the lowest-order G-matrix theory with the CON choice.

One of great successes of the G-matrix approach is that
nucleon-nucleus and nucleus-nucleus scattering observables are nicely 
reproduced with the complex G-matrix folding potentials derived from
free-space $N\!N$ interactions.
In ref.\cite{FSY},
it was shown clearly that the TNR effect appeared 
in angular distributions of $^{16}$O+$^{16}$O elastic scattering
($E/A$=70 MeV), $etc$. Their analysis is used here
to determine the coupling constants in MPP:
The G-matrix folding potentials including MPP contributions
are used to analyze the $^{16}$O+$^{16}$O scattering, and then 
the strengths of MPP are adjusted so as to reproduce the experimental data.

Many attempts have been made to extract some information on the 
incompressibility $K$ of high-density matter formed in high-energy 
central heavy-ion collisions. In many cases, however,
the results for the EoS still remain inconclusive.
On the other hand, it was pointed out that folding-model analyses
of high-precision nuclear scattering data can be used as an independent
method to determine the value of $K$, which was demonstrated with use
of density-dependent interactions~\cite{Khoa95,Khoa97}.
Our approach can be considered as a development from theirs:
The MPP contributions determined by analyses with the G-matrix 
folding model are included in constructing the EoS of neutron-star matter,
which is expected to result in a stiff EoS, enough to give 
the observed neutron-star mass. 

The important feature of our MPP is that 
it works universally not only in $N\!N\!N$ states but also
$Y\!N\!N$, $Y\!Y\!N$ and $Y\!Y\!Y\!$ states.
It will be shown in our future work that inclusions of 
MPP's in hypernuclear calculations lead to reasonable results.

\section{Multi-pomeron potential}
We introduce the universal TBR 
consistently with the ESC modeling, assuming that
the dominant mechanism is triple and quartic pomeron exchange.
For the N-tuple pomeron vertex generally,
we take the Lagrangian
\begin{eqnarray}
 {\cal L}_N = g_P^{(N)} {\cal M}^{4-N} \sigma_P^N(x)/N!
\label{eq:tbf.1}
\end{eqnarray}
The N-body local potential by pomeron exchange is
\begin{eqnarray}
 && V({\bf x}'_1, ..., {\bf x}'_N; {\bf x}_1, ... , {\bf x}_N) \equiv
 V({\bf x}_1, ..., {\bf x}_N)\ \Pi_{i=1}^N \delta({\bf x}'_i-{\bf x}_i), 
 \nonumber
\end{eqnarray}
\begin{eqnarray}
 && V({\bf x}_1, ..., {\bf x}_N) = g_P^{(N)} g_P^N\ \left\{
\int\frac{d^3k_i}{(2\pi)^3} e^{-i{\bf k}_i\cdot{\bf x}_i}\right\}
\nonumber\\ && \times (2\pi)^3\delta(\sum_{i=1}^N {\bf k}_i)
\Pi_{i=1}^N \left[\exp\left(-{\bf k}_i^2\right)\right]{\cal M}^{4-3N},
\end{eqnarray}
where the (low-energy) pomeron propagator is the same as used in the
two-body pomeron potential.
Since the pomeron is an SU(3)-singlet, MPP's work universally
among baryons. 

The effective two-body potential in a baryonic medium is obtained
by integrating over the coordinates ${\bf x}_3,..., {\bf x}_N$. 
This gives 
\begin{eqnarray}
&& V_{eff}^{(N)}({\bf x}_1,{\bf x}_2) 
 = \rho_{NM}^{N-2} 
 \int\!\! d^3\!x_3 ... \int\!\! d^3\!x_N\ 
 V({\bf x}_1,{\bf x}_2, ..., {\bf x}_N)
 \nonumber
 \\
&& 
=g_P^{(N)} g_P^N\frac{\rho_{NM}^{N-2}}{{\cal M}^{3N-4}}
 \frac{1}{\pi\sqrt{\pi}} \left(\frac{m_P}{\sqrt{2}}\right)^3
 \exp\left(-\frac{1}{2}m_P^2 r_{12}^2\right),
\label{eq:tbf.3}
\end{eqnarray}
$\rho_{NM}$ being a nuclear-matter density.
%
We restrict ourselves here to the triple and quartic pomeron couplings.

Values of MPP strengths $g_P^{(3)}$ can be estimated
from the experimental cross sections of the process $pp \rightarrow pX$
(diffractive production of showers of particles) 
at very high energies~\cite{Kai74}:
The estimated values are $g_P^{(3)}/g_P= 0.15 \sim 0.20$.
Using the value of $g_P=3.67$ used in ESC, we have
$$g_P^{(3)}= 1.95 \sim 2.6 \ .$$
In the Reggeon field theory~\cite{Bron77},
the value of $g_P^{(4)}$ can be estimated as
$$
g_P^{(4)} = (8.8 \sim 60)\, (g_P^{(3)})^2 = 33 \sim 228 \ . 
$$


\section{MPP in nucleus-nucleus scattering}
In \cite{FSY} the complex G-matrix interaction was derived from 
the ESC model including TNR, and nucleus-nucleus elastic scatterings were analyzed
with the use of double-folding potentials. In this work, the TNR effect was assumed 
to come from the medium effect on the vector-meson masses differently 
from the present MPP model. When a scattering energy is high enough, 
the frozen-density approximation gives a good prescription.
In this approximation, G-matrices at about two times of normal density 
contribute to folding potentials.
Then, the TNR effects were shown to appear clearly in angular distributions
in the case of $^{16}$O$+^{16}$O elastic scattering at $E/A=70$ MeV,
and also in the cases of $^{16}$O elastic scattering by the $^{12}$C, $^{28}$Si
and $^{40}$Ca at $E/A=93.9$ MeV, and $^{12}$C$+^{12}$C elastic scattering 
at $E/A=135$ MeV.
Here, the same analyses are performed so that the MPP strengths 
$g_P^{(3)}$ and $g_P^{(4)}$ are determined to reproduce the experimental data 
with the use of the G-matrix folding potential derived from ESC including MPP.

On the other hand, the nuclear saturation property cannot
be reproduced only by adding MPP to ESC. 
Then, we introduce also a TNA part phenomenologically as
a density-dependent two-body interaction
\begin{eqnarray}
V_{TNA}(r;\rho)= V^0_{TNA}\, \exp(-(r/2.0)^2)\, \rho\, 
\exp(-4.0\rho)\ .
\end{eqnarray}
Here, because the functional form is not determined within 
our analysis, it is fixed to be similar to the TNA part 
given in \cite{Panda81}. Only the strength $V^0_{TNA}$ is treated as
an adjustable parameter.

This TNA is assumed to work only in even states:
An odd-state TNA contributes to the $^{16}$O$+^{16}$O potential
at $E/A=70$ MeV more than the nuclear-matter energy, because
G-matrices in the former are with higher momenta than those
in the latter. Therefore, they cannot be reproduced consistently,
for instance, with use of a Wigner-type TNA.
Also the Fujita-Miyazawa TNA used in \cite{FSY} is specified
by dominant contributions from even-state components.

On the basis of G-matrix calculations, strengths of
the TNR part ($g_P^{(3)}$ and $g_P^{(4)}$) and the TNA part ($V^0_{TNA}$)
are determined so as to reproduce the $^{16}$O$+^{16}$O angular distribution
$E/A=70$ MeV and the minimum value $\sim -16$ MeV of the energy per nucleon
at normal density in symmetric matter.
Because the ratio of $g_P^{(3)}$ and $g_P^{(4)}$ cannot be 
determined in our analysis, it is taken adequately referring the above 
estimation based on \cite{Kai74,Bron77}.
Then, we have ($g_P^{(3)}$, $g_P^{(4)}$)= (2.34, 30.0) and $V^0_{TNA}=-36$ MeV.
This set is called as MP1a. 
In order to see a role of $g_P^{(4)}$ especially in high-density region,
we take here another set ($g_P^{(3)}$, $g_P^{(4)}$)= (2.94, 0.0) with
the same value of $V^0_{TNA}$. This set without $g_P^{(4)}$ is called MP2a.

\begin{figure}[htb!] 
\centering 
\includegraphics[width=6cm]{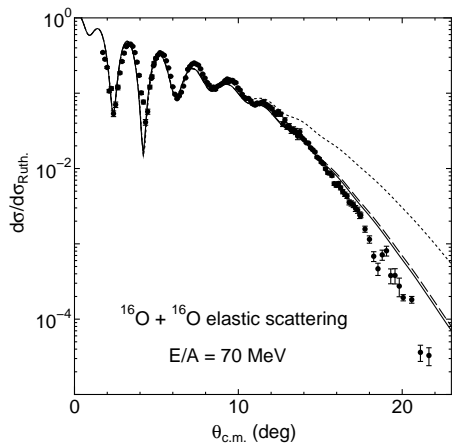} 
\caption{\small \label{xsO16O16}
Differential cross sections for $^{16}$O+$^{16}$O elastic scattering at
$E/A=70$ MeV calculated with the G-matrix folding potentials.
Solid  and dashed curves are for MP1a and MP2a, respectively.
Dotted curve is for MP0. 
}
\end{figure} 

In Fig.\ref{xsO16O16}, the calculated results of the differential cross sections 
for the $^{16}$O+$^{16}$O elastic scattering at $E/A=70$ MeV are compared with
the experimental data~\cite{Nuoffer}. 
Here, the dotted curve is obtained with the G-matrix folding potential derived 
from MP0 meaning ESC without MPP, which deviates substantially from the data.
Solid  and dashed curves are for MP1a and MP2a, respectively,
which are found to fit the data nicely.
Though reduction factors are often multiplied on the imaginary parts
in the folding model analyses~\cite{FSY},
such a reduction factor is not used in the present analysis.
The necessity to include the quartic pomeron coupling has to appear in 
the difference between results for MP1a and MP2a, but it cannot be found 
in the present analyses for nucleus-nucleus scattering.

\section{Equation of state}
\begin{figure}[htb!] 
\centering 
\includegraphics[width=6cm]{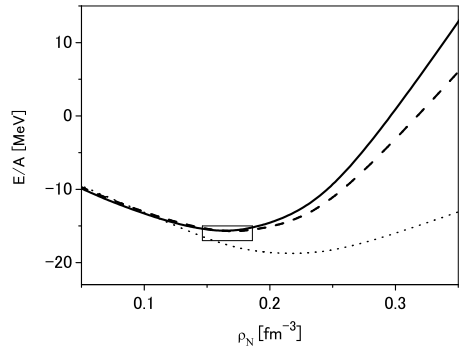} 
\caption{\small \label{saturation1}
Energy per particle of symmetric matter 
as a function of $\rho_N$.
Solid and dashed curves are for MP1a and MP2a), respectively.
Dotted curve is for MP0.
The box shows the empirical value.
}
\end{figure}

In Fig.\ref{saturation1}, 
energies per nucleon in symmetric nuclear matter
are drawn as a function of nucleon density $\rho_N$. 
The box in the figure show the area where nuclear saturation is 
expected to occur empirically.
The dotted curve is obtained for MP0. 
Then, the saturation density and the minimum energy in symmetric matter 
are found to be deviated substantially from the box.
Solid and dashed curves are for MP1a and MP2a, respectively.
As seen clearly, saturation densities and minimum values in 
these cases are close nicely to the empirical value shown by the box.
Then, obtained values of incompressibility $K$ are 280 MeV and 270 MeV
at minimum point in the cases of including MP1a and MP2a, respectively.
which are comparable to the empirical value
$240 \pm 20$ MeV~\cite{Colo,Sagawa}.


\begin{figure}[htb!] 
\centering 
\includegraphics[width=6cm]{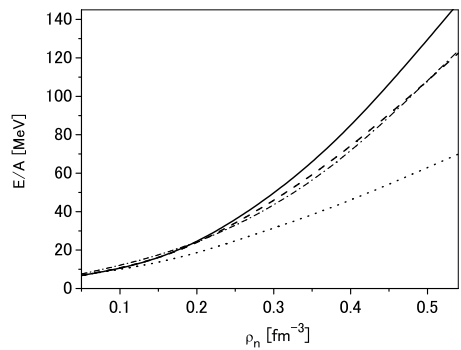} 
\caption{\small \label{saturation2}
Energy per particle of neutron matter 
as a function of $\rho_n$. 
Solid and dashed curves are for MP1a and MP2a, respectively.
Dotted curve is for MP0.
Dot-dashed curve is for UIX~\cite{Gandolfi12}.
}
\end{figure}

In Fig.\ref{saturation2}, 
we show energy curves of neutron matter, namely binding energy
per neutron as a function of neutron density $\rho_n$.
Solid and dashed curves are for MP1a and MP2a, respectively.
Dotted curve is for MP0.
The difference between the energy curves for neutron matter and
symmetric matter gives the symmetry energy $E_{sym}(\rho)$, and
its slope parameter is defined by 
$L=3\rho_0 \left[\frac{\partial E_{sym}(\rho)}{\partial \rho}\right]$.
The values of $E_{sym}$ at normal density are 32.5, 33.4 and 33.5 MeV
in the cases of MP0, MP1a and MP2a, respectively, and
the values of $L$ are 71, 72 and 73 MeV correspondingly.
These values are in nice agreement to the values 
$E_{sym}=32.5\pm0.5$ MeV and $L=70\pm15$ MeV determined 
recently on the basis of experimental data~\cite{Yoshida}.
Thus, the nuclear saturation property derived from MP1a or MP2a 
is quite reasonable in comparison with the empirical values,
as seen in obtained values of $K$, $E_{sym}$ and $L$.


\section{Neutron star mass}

Recently, 
the important roles of the three-neutron $(3n)$ repulsion are studied
for the EoS of neutron matter and the neutron-star mass~\cite{Gandolfi12}.
Among their interactions, the stiffest EoS is given by the UIX model
(the Argonne AV8' model combined with the Urbana IX).
This EoS is stiff enough to explain the large neutron-star mass
$1.97 M_{solar}$~\cite{Demorest10}.
%
Their result for the UIX model 
is shown by the dot-dashed curve in Fig.\ref{saturation2}.
Our results for MP1a (MP2a) turns out to be stiffer than
(comparable to) theirs. 

Using the EoS of pure neuron matter, we solve the 
Tolmann-Oppenheimer-Volkoff (TOV) equation for the hydrostatic structure
of a spherical non-rotating star, and obtain the mass and 
radius of neutron stars. 
From the G-matrix calculations, we obtain the energy per neutron 
as a function of density $\rho$.
According to \cite{Gandolfi12}, the energy curve is parameterized as
\begin{eqnarray}
E(\rho)/A = a \left(\frac{\rho}{\rho_0}\right)^\alpha
        + b \left(\frac{\rho}{\rho_0}\right)^\beta \ .
\end{eqnarray}
The fitting parameters are given in Table \ref{Param}
in the cases of MP0, MP1a and MP2a.
These EoS's are used for $\rho > \rho_{crust} = \rho_0/10$.
Below $\rho_{crust}$ we use the EoS of the crust obtained
in \cite{Baym1,Baym2}.
\begin{table}[ht]
\centering 
\setlength{\textwidth}{50mm} 
\caption{Fitting parameters for the neutron matter EoS,
$a$ and $b$ being in MeV.
}
\label{Param}
 \begin{tabular}{ccccc}
 \hline\hline
& $a$ & $\alpha$ & $b$ & $\beta$ \\ 
 \hline
MP0  & 10.8 & 0.442 &  5.31 &  1.88  \\
MP1a & 11.4 & 0.481 &  7.70 &  2.41  \\
MP2a & 6.79 & 0.192 &  11.6 &  1.93  \\
\hline\hline 
\end{tabular}
\end{table}

The EoS's for MP1a and MP2a violate causality and predict 
sound speeds over the speed of light above a critical density.
Then, we adopt the approximation where the EoS is replaced by
the causal EoS above this density
in the same way as the treatment in \cite{Gandolfi12}.

\begin{figure}[ht] 
\begin{center} 
\includegraphics*[width=6cm]{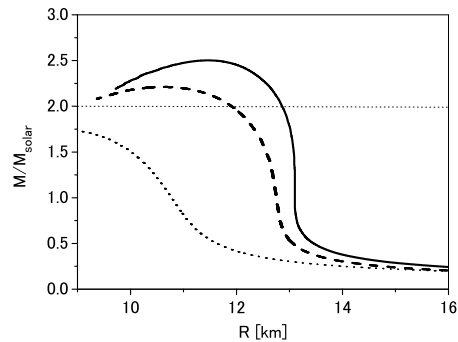} 
\caption{\small \label{starmass1}
Neutron-star masses as a function of the radius $R$.
Solid and dashed curves are for MP1a and MP2a, respectively.
Dotted curve is for MP0.
}
\end{center}
\end{figure}

\begin{figure}[ht] 
\begin{center} 
\includegraphics*[width=6cm]{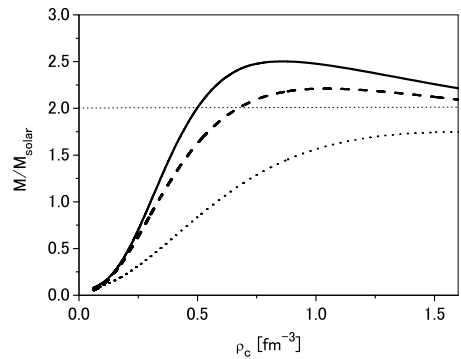} 
\caption{\small \label{starmass2}
Neutron-star masses as a function of the central density $\rho_c$.
Also see the caption of Fig.\ref{starmass1}.
}
\end{center}
\end{figure}

In Fig.\ref{starmass1} and Fig.\ref{starmass2}, the calculated
star masses are given as a function of the radius $R$ and
the central density $\rho_c$, respectively.
Solid and dashed curves are for MP1a and MP2a, respectively.
Dotted curves are for MP0.
The EoS's in the cases of including MPP contributions are found to be 
stiff enough to give maximum masses larger than the observed mass 
$M_{obs}=1.97 M_{solar}$ for J1614-2230.  
The maximum masses for MP1a are substantially 
larger than that for MP2a.
The difference is due to the quartic-pomeron exchange term
included in the former. The strengths of the effective two-body 
interaction derived from triple- and quartic-pomeron exchanges 
are proportional to matter density $\rho$ and $\rho^2$, respectively.
Then, the contribution of the latter become sizeable
in the high-density region, which makes the maximum mass large.
However, the difference between MP1a and MP2a contributions
cannot be seen in Fig.\ref{xsO16O16}, which means that
the quartic-pomeron exchange contribution demonstrated in
neutron-star masses cannot be found by analyses of 
nucleus-nucleus scattering data.



\section{Conclusion}
In order to explain the observed maximum mass of neutron stars,
three-body repulsions are considered to work universally among
three baryons.
The multi-pomeron potential (MPP) is introduced as a model for such a
universal three-body repulsion.
Furthermore, the three-nucleon attraction (TNA) is added phenomenologically
so as to reproduce the nuclear saturation property precisely.
The strengths of MPP and TNA can be determined by fitting the observed 
angular distribution of $^{16}$O+$^{16}$O elastic scattering at $E/A=70$ MeV
with use of the G-matrix folding potential derived from ESC+MPP+TNA
and adjusting to reproduce the minimum value $\sim -16$ MeV of the
energy per nucleon at normal density in symmetric nuclear matter. 
Then, the empirical values of $K$, $E_{sym}$ 
and $L$ are reproduced reasonably.
The EoS of neutron-star matter obtained from ESC+MPP+TNA is stiff 
enough to give the large neutron-star mass over $1.97 M_{solar}$.
It should be noted that our stiff EoS is determined with use
of the experimental data for nucleus-nucleus scattering and
nuclear saturation properties.
Our MPP contributions exist universally in every baryonic system.
It is our future subject to show that our stiff EoS with 
universal MPP repulsions is free from the softening effect
induced by hyperon mixing to neutron-star matter.
Such a work will be complementary to the analyses based on the
relativistic mean field models \cite{Weiss,Bednarek,Jiang}
for a massive neutron star including hyprons.

\section*{Acknowledgments}
We wish to thank Prof. T. Uesaka for giving us
the motivation of this work.


\begin{thebibliography}{99}

\bibitem{Demorest10}
P.B. Cemorest, T. Pennucci, S.M. Ransom, M.S.E. Roberts, and J.W. Hessels,
Nature (London) {\bf 467}, 1081 (2010). 

\bibitem{NYT}
S. Nishizaki, Y. Yamamoto, and T. Takatsuka,
Prog. Theor. Phys. {\bf105}, 607 (2001); {\bf 108}, 703 (2002).

\bibitem{ESC08} 
Th.A. Rijken, M.M. Nagels, and Y. Yamamoto, 
Prog. Theor. Phys. Suppl. {\bf 185}, 14 (2010). 

\bibitem{ESC08c} 
Th.A. Rijken, M.M. Nagels, and Y. Yamamoto,
in {\it Proceedings of the International Workshop on Strangeness
Nuclear Physics}, Neyagawa 2012,
Genshikaku Kenkyu {\bf 57}, Suppl.3, 6 (2013). 

\bibitem{Baldo02}
M. Baldo, A. Fiasconaro, H.Q. Song, G. Giansiracusa, and U. Lombardo,
Phys. Rev. C{\bf 65}, 017303 (2001). 

\bibitem{FSY}
T. Furumoto, Y. Sakuragi, and Y. Yamamoto, 
Phys. Rev. C{\bf 79}, 011601(R) (2009); C{\bf 80}, 044614 (2009).

\bibitem{Khoa95}
D.T. Khoa, et al.,
Phys. Rev. Lett. {\bf 74}, 34 (1995).

\bibitem{Khoa97}
D.T. Khoa, G.R. Satchler, and W.von Oertzen,
Phys. Rev. C{\bf 56}, 954 (1997).

\bibitem{Kai74}
A.B. Kaidalov and K.A. Ter-Materosyan, 
Nucl. Phys. {\bf 74}, 471 (1974).

\bibitem{Bron77}
J.B. Bronzan and R.L. Sugar, 
Phys. Revs. D{\bf 16}, 466 (1977).

\bibitem{Nuoffer}
F. Nuoffer, et al., 
Nuovo Cimento A{\bf 111}, 971 (1998).

\bibitem{Panda81}
I.E. Lagaris and V.R. Pandharipande,
Nucl. Phys. A{\bf 359}, 349 (1981).

\bibitem{Colo}
G. Col\`{o}, N. Van Giai, J. Meyer, K. Bennaceur, and P. Bonche,
Phys. Rev. C{\bf 70}, 024307 (2004).

\bibitem{Sagawa}
Li-Gang Cao, H. Sagawa, and G. Colo,
Phys. ReV. C{\bf 86}, 054313 (2012).

\bibitem{Yoshida}
P. M\"{o}ller, W.D. Myers, H. Sagawa and S. Yoshida,
Phys. Rev. Lett. {\bf 108}, 052501 (2012).

\bibitem{Gandolfi12}
S. Gandolfi, J. Carlson, and Sanjjay Reddy,
Phys. Rev. C{\bf 85}, 032801(R) (2012). 

\bibitem{Baym1}
G. Baym, A. Bethe, and C. Pethick,
Nucl. Phys. A{\bf 175}, 225 (1971).

\bibitem{Baym2}
G. Baym, C.J. Pethick, and P. Sutherland,
Astrophys. J.{\bf 170}, 299 (1971).


\bibitem{Weiss}
S. Weissenborn, D. Chatterjee, and J. Schaffner-Bielich,
Nucl. Phys. A{\bf 881}, 62 (2012).

\bibitem{Bednarek}
I. Bednarek, P. Haensel, J.L. Zdunik, M. Bejger, and R. Manka,
Astronomy \& Astrophysics A{\bf 157}, 543 (2012).

\bibitem{Jiang}
Wei-Zhou Jiang, Bao-An Li, and Lie-Wen Chen,
Astrophys. J.{\bf 756}, 56 (2012).




\end{thebibliography}
\end{document}